\documentclass[11pt]{article}

\usepackage[preprint]{acl}

\usepackage{times}
\usepackage{latexsym}

\usepackage{graphicx}
\usepackage{multirow}
\usepackage{subcaption} 
\usepackage{tabularx} 
\usepackage{booktabs} 
\usepackage{geometry} 
\usepackage{amsmath}

\usepackage[normalem]{ulem}
\useunder{\uline}{\ul}{}

\usepackage{colortbl}
\usepackage{booktabs}

\definecolor{lamdocolor}{RGB}{220,20,60}

\definecolor{linhnguyencolor}{RGB}{20, 60, 220}

\definecolor{customdarkorange}{RGB}{255,146,69}
\definecolor{customdarkblue}{RGB}{121,160,255}

\usepackage[T1]{fontenc}

\usepackage[utf8]{inputenc}

\usepackage{microtype}

\usepackage{inconsolata}

\usepackage{graphicx}

\title{CASPER: Concept-integrated Sparse Representation for Scientific Retrieval}

\author{
    Lam Thanh Do\textsuperscript{1}\thanks{Equal contribution}, 
    Linh Van Nguyen\textsuperscript{2}\footnotemark[1], 
    Jiayu Li\textsuperscript{1}, 
    David Fu\textsuperscript{1}, 
    Kevin Chen-Chuan Chang\textsuperscript{1} \\
    \textsuperscript{1}University of Illinois Urbana-Champaign \quad
    \textsuperscript{2}Aalto University \\
    \texttt{\{lamdo, jiayul11, jiahaof4, kcchang\}@illinois.edu} \\
    \texttt{linh.10.nguyen@aalto.fi}
}

\begin{document}
\maketitle
\begin{abstract}

Identifying relevant research concepts is crucial for effective scientific search. However, primary sparse retrieval methods often lack concept-aware representations. To address this, we propose CASPER, a sparse retrieval model for scientific search that utilizes both tokens and keyphrases as representation units (i.e., dimensions in the sparse embedding space). This enables CASPER to represent queries and documents via research concepts and match them at both granular and conceptual levels. Furthermore, we construct training data by leveraging abundant scholarly references (including titles, citation contexts, author-assigned keyphrases, and co-citations), which capture how research concepts are expressed in diverse settings. Empirically, CASPER outperforms strong dense and sparse retrieval baselines across eight scientific retrieval benchmarks. We also explore the effectiveness-efficiency trade-off via representation pruning and demonstrate CASPER's interpretability by showing that it can serve as an effective and efficient keyphrase generation model\footnote{Code: \url{https://github.com/louisdo/CASPER}}.

\end{abstract}

\section{Introduction}
\label{sec:introduction}

Science progress is accelerating at an unprecedented rate, with millions of research articles published yearly \cite{fortunato2018science}. This surge in scientific output makes it increasingly difficult for researchers to catch up with the latest developments \cite{huettemann2025designing}. As a result, there is a pressing need for robust information retrieval systems that can efficiently navigate and filter the vast landscape of scientific literature.


To address this challenge, the community has largely turned to Dense Retrieval (DR) models \cite{cohan-etal-2020-specter, singh-etal-2023-scirepeval, mysore-etal-2022-multi, kang-etal-2024-taxonomy, kang2025improving}, which encode documents into continuous vector spaces to capture semantic nuances. However, Learned Sparse Retrieval (LSR) has recently emerged as an attractive alternative. Unlike DR models, LSR models encode text into sparse lexical vectors, where each term (word or subword) serves as a representation unit, i.e., a dimension in the embedding space. The advantages of LSR over dense retrieval include 1) efficiency due to its compatibility with inverted indexes and 2) interpretability since each dimension in the representation space corresponds to a term. Realizing the full potential of LSR in this context of scientific document search, however, requires overcoming three challenges.


\begin{table}[]
\centering

\resizebox{0.9\columnwidth}{!}{

\begin{tabular}{@{}ll@{}}
\toprule
\textbf{Query}  & deep transfer learning in neural networks                                                                                                                                                                                                                 \\ \midrule
\textbf{SPLADE} & \begin{tabular}[c]{@{}l@{}}(deep, 2.55), (transfer, 2.27), (neural, 1.73), (learning, 1.56), \\ (brain, 1.37), (networks, 1.33), (learn, 1.26), (network, 1.15), \\ (transferred, 1.14), (college, 1.03), (,, 0.69), (goal, 0.61), ...\end{tabular}       \\ \midrule
\textbf{CASPER} & \begin{tabular}[c]{@{}l@{}}(deep learning, 2.26), (deep, 1.43), (knowledge transfer, 1.36), \\ (transfer learning, 1.25), (soap, 1.24), (neural network, 1.15), ..., \\ (domain adaptation, 0.55), ..., (artificial intelligence, 0.41), ...\end{tabular} \\ \bottomrule
\end{tabular}

}
\caption{SPLADE and CASPER representations for a hypothetical query.}
\label{tab:representation_example}
\vspace{-0.5cm}
\end{table}


The \textbf{first challenge} is \textit{the lack of conceptual representation units}. Primary sparse retrieval approaches \cite{bai2020sparterm, zhao2020sparta, formal2021splade, formal2021spladev2} utilize BERT tokens as representation units. However, it has been pointed out that searching for scientific article heavily involves identifying research concepts to include for querying \cite{bramer2018systematic}. Although tokens allow for matching of granular details, the lack of conceptual representation unit hinder capturing research concepts, potentially causing the model to miss relevant documents. For instance, consider the example query in Table \ref{tab:representation_example}. Other than documents that discuss ``transfer learning'', it is expected that those discussing relevant concepts, such as ``knowledge transfer'' and ``domain adaptation'', are also retrieved. While SPLADE \cite{formal2021splade, formal2021spladev2}, a prominent sparse model, expands the query representation beyond the present tokens to include related ones. Consequently, SPLADE’s representation fails to capture the conceptual relationship and miss relevant articles.

\begin{figure}
    \centering
    \includegraphics[width=\columnwidth]{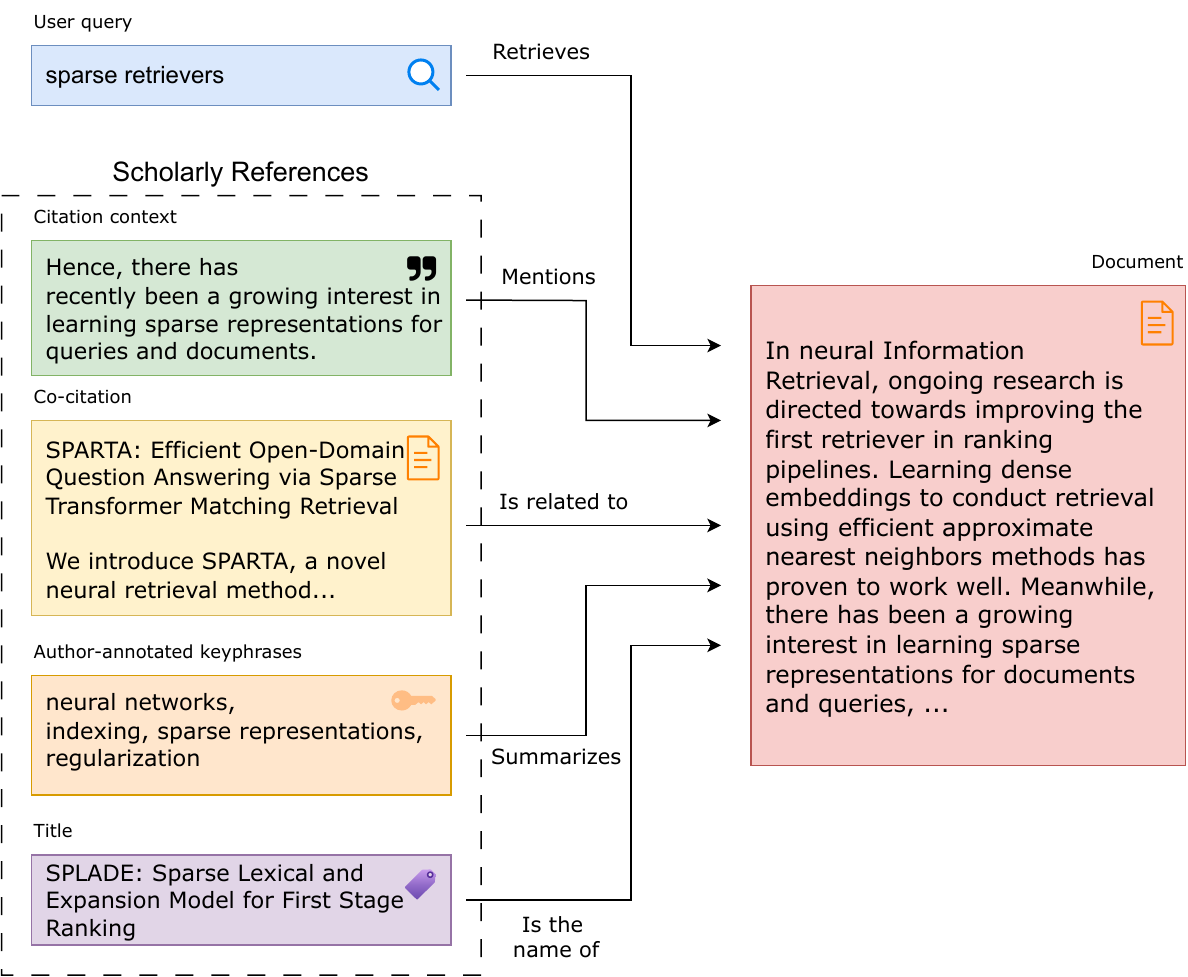}

    \caption{Example of scholarly references}
    \label{fig:ExampleScholarlyReferences}

    \vspace{-0.5cm}
\end{figure}

Our key idea for addressing this challenge is to \textit{learn concept-integrated representation}. In particular, we propose to treat research concepts as representation units in addition to tokens, allowing matching both token and concept-level details.


The \textbf{second challenge} is that it is unclear regarding the criterion to \textit{determine conceptual representation units}. We propose \textit{selecting common and comprehensive keyphrase vocabulary as concept units}. Keyphrases are typically used to describe research concepts and therefore are a natural choice for concept units. The selected keyphrases should be common, widely present in queries and documents, to ensure sufficient instances for each dimension for the model to learn. Furthermore, the keyphrase vocabulary should be comprehensive, covering the diverse research concepts circulating in the literature, enabling each document to be represented with these keyphrases. For instance, a vocabulary limited to Computer Science terminology cannot effectively represent Biology articles.



The \textbf{third challenge} is \textit{limited supervision signals}. Existing data sources do not support learning concept-aware representation. Previous work utilizes pairs of articles, formed by either citation links \cite{cohan-etal-2020-specter, singh-etal-2023-scirepeval} or co-citations \cite{mysore-etal-2022-multi}. However, besides the potential inaccuracies of these signals (as in the case of citation links \cite{mysore-etal-2022-multi}), relying solely on them is insufficient. More specifically, they mainly indicate related concepts mentioned in \textit{other works}, but do not directly reveal the concepts expressed in the given text itself.


User queries and interaction are another potential source of supervision, with an example being SciRepEval Search \cite{singh-etal-2023-scirepeval}. However, such signals are typically proprietary and difficult to obtain. Moreover, they are also insufficient. Our analysis reveals that a significant portion of queries in SciRepEval Search closely match paper titles. In particular, 44\% of query–document pairs involve queries that appear as substrings in the document title or vice versa, and 53\% contain all query terms within the document title. This pattern indicates predominantly known-item rather than exploratory, concept-driven searches. It is noteworthy that this reflects the limitations of current scholarly search engines, which primarily support known-item retrieval \cite{gusenbauer2021every}, rather than scientific document search being a simple task.


Our key idea for solving this challenge is \textit{leveraging diverse scholarly references}. We refer to scholarly references as signals within the literature to connect and describe academic articles \textit{at the concept level}. We specifically utilize four sources namely \textit{titles}, \textit{citation contexts}, \textit{author-assigned keyphrases}, and \textit{co-citation networks} as (pseudo) queries to retrieve the underlying referenced documents. These signals are embedded within the scientific literature and are therefore ``free'' to obtain. In addition, they capture research concepts of papers expressed in different settings and therefore are rich sources of supervision. In Figure \ref{fig:ExampleScholarlyReferences}, we provide an example of these sources of supervision.

We summarize our contributions as follows. \textbf{Firstly}, we propose \textbf{CASPER}, a sparse retrieval model that utilizes both tokens and keyphrases as representation units. \textbf{Secondly}, we introduce a framework to construct large-scale scientific Information Retrieval (IR) training data by mining scholarly references, which we name \textbf{FRIEREN}. \textbf{Finally}, we extensively evaluate CASPER on eight scientific retrieval benchmarks, where it outperforms strong dense and sparse baselines. We also explore the effectiveness-efficiency trade-off via representation pruning and demonstrate CASPER's interpretability by showing its utility as a robust keyphrase generation model.



\section{Related work}

\begin{figure*}[ht]
    \centering
    \includegraphics[width=0.9\textwidth]{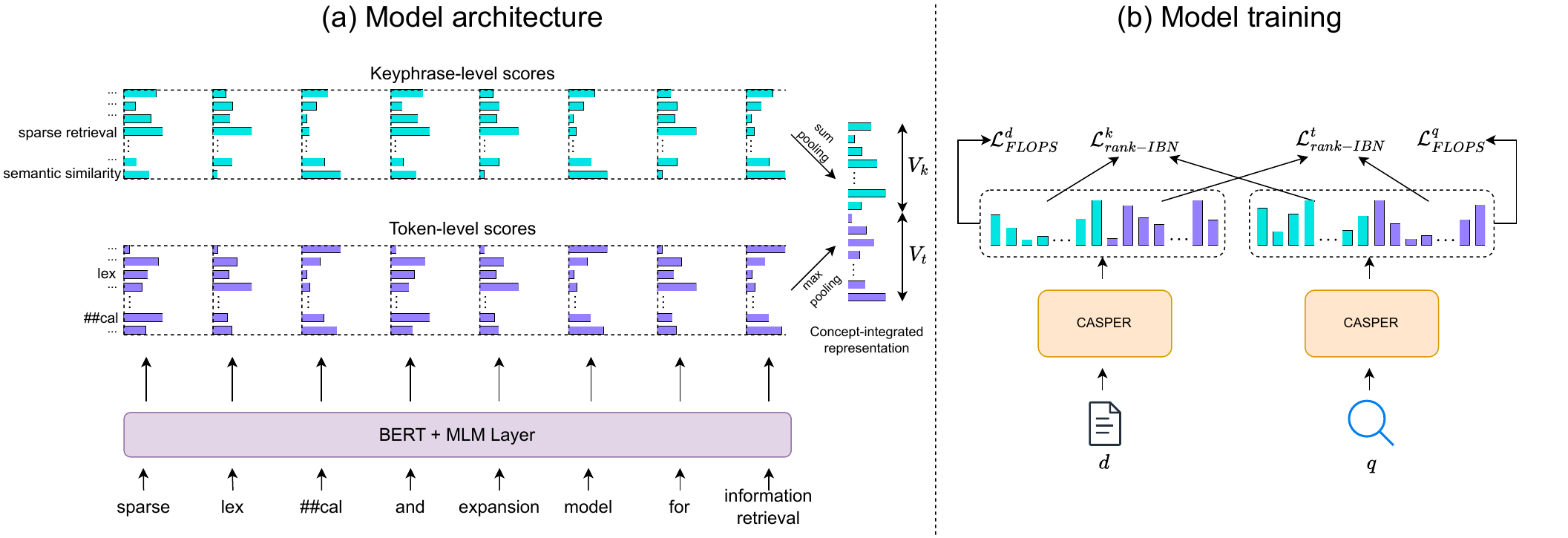}
    \caption{Overview of CASPER, our proposed method.}
    \label{fig:CASPEROverview}
\end{figure*}


\textbf{Learned Sparse Retrieval}. Sparse retrieval models encode inputs as term-based vectors. Traditional methods, such as TF-IDF \cite{salton1988term} and BM25 \cite{robertson2009probabilistic}, generate bag-of-words representations from document and corpus statistics. However, these approaches suffer from \textit{vocabulary mismatching} problem. To address this, primary methods \cite{bai2020sparterm, dai2019context, zhao2020sparta, formal2021splade, formal2021spladev2} leverage pretrained language models to learn contextualized sparse representations. By assigning importance scores to tokens from the entire vocabulary, rather than limiting them to those present in the input, these methods effectively mitigate vocabulary mismatch.

Most sparse retrieval models represent text in BERT's vocabulary. Expanded-SPLADE \cite{dudek2023learning} attempts to replace this vocabulary with a customized vocabulary of the 300k most frequent English unigrams. DyVo \cite{nguyen2024dyvo} enhances BERT’s vocabulary by incorporating entities retrieved from a knowledge base. Our proposed method, is more related to DyVo. However, while CASPER learns an end-to-end representation integrating both tokens and keyphrases as representation units, DyVo augments a token-based representation with a frozen entity retrieval system, meaning the association between text and (implicit) entities is fixed and not learned from training data.


\noindent \textbf{Retrieval in the Scientific Domain.} Information retrieval in the scientific domain has attracted increasing attention in recent years. Early work \cite{el2011beyond} proposed recommendation methods for scientific articles using features such as citation links and shared authorship. Motivated by the multifaceted nature of scientific texts, aspect-based representations have been explored \cite{jain2018learning, neves2019evaluation, chan2018solvent, kobayashi2018citation, mysore2021multi}. Another line of research emphasizes the importance of research concepts-aware representation \cite{kang-etal-2024-taxonomy, kang2025improving}, to which CASPER belongs. Building upon the foundations laid by prior studies, our work introduces two key distinctions: 1) we propose a sparse retrieval model, whereas existing work focuses on dense retrieval; 2) CASPER is trained end-to-end and learns to discover concepts directly from data, while prior methods rely on external systems (such as LLMs) for their concept representation. 

Regarding the supervision signals to train scientific IR models, existing research have introduced citation-based document representations \cite{cohan-etal-2020-specter, singh-etal-2023-scirepeval}, where models are trained on pairs of papers linked by a citation. However, citation links can be a noisy proxy for semantic relevance, therefore \citet{mysore2021multi} propose co-citation, a more accurate source of supervision. Another source is user queries and interaction data, exemplified by SciRepEval Search \cite{singh-etal-2023-scirepeval}. Our work leverages \textit{scholarly references}, which we define as signals within the literature that connect and describe academic articles at the concept level. These signals are not only easy to obtain, but are also rich as they capture research concepts of articles in different settings.


\noindent \textbf{Keyphrase Generation}. Keyphrase generation involves producing phrases that capture the core concepts of a text. Prior work typically formulated as a supervised sequence-to-sequence learning task \cite{meng-etal-2017-deep}. Although keyphrases have proven useful for scientific retrieval \cite{boudin-etal-2020-keyphrase}, focused research on keyphrase generation to enhance retrieval remains limited. Notable exceptions explore the use of keyphrases for query and document expansion \cite{wu2022fast, do2025eru}. Our work lies in the intersection of keyphrase generation and information retrieval. Specifically, CASPER is a sparse retrieval model that explicitly represents documents and queries using keyphrases, thereby integrating the capabilities of a keyphrase generation model within a retrieval framework.


\section{Preliminary: SPLADE}


\textbf{Architecture}. SPLADE \cite{formal2021splade, formal2021spladev2} is a sparse represetation method that predicts term importances for query/document in the BERT \cite{devlin-etal-2019-bert} token vocabulary space $V_t$. In particular, given an input document $d$, let $w_{ij}^d$ denotes the importance of token $j \in V_t$ given token $i \in d$, predicted by the Masked Language Modeling (MLM) layer. The importance $w_j^d$ of $j$ given $d$ is computed with max pooling \begin{equation}
    w_j^d = \max_{i \in d} \log (1 + \text{ReLU}(w_{ij}^d))
\end{equation}

\noindent \textbf{Training}. Training SPLADE requires a collection of triplets $(q_i, d_i^{+}, d_i^{-})$, where $q_i$, $d_i^{+}$, $d_i^{-}$ respectively denotes query, positive document and hard negative document; the set of in-batch negatives is denoted as $\{d_{i,j}^-\}_j$. SPLADE is trained by optimizing a contrastive loss, with both in batch and hard negatives; along with sparsity regularization
\begin{equation}
\label{eq:SPLADE_training_objective}
    \mathcal{L} = \mathcal{L}_{rank-IBN} + \lambda_d \mathcal{L}_{FLOPS}^d + \lambda_q \mathcal{L}_{FLOPS}^q
\end{equation}

\begin{equation}
\resizebox{\columnwidth}{!}{

$\mathcal{L}_{rank-IBN} = -\log \frac{e^{s(q_i,d_i^+)}}{e^{s(q_i,d_i^+)} + e^{s(q_i,d_i^-)} + \sum_j e^{s(q_i,d_{i,j}^-)}}$

}
\label{eq:SPLADE_ranking_loss}
\end{equation}
where $s(q,d) = \sum_{j \in V_t} w_j^d \ w_j^q$ is the similarity between query $q$ and document $d$, defined by the dot product of their representation. $\mathcal{L}_{FLOPS}$ is the FLOPS regularizer to enforce sparsity, first introduced in \cite{paria2020minimizing}. For further details on SPLADE, we encourage readers to refer to the original papers \cite{formal2021splade, formal2021spladev2}.

\section{CASPER}

In this section, we detail CASPER, a \textbf{C}oncept-integr\textbf{A}ted \textbf{SP}ars\textbf{E} \textbf{R}epresentation model for scientific search; an overview is presented in Figure \ref{fig:CASPEROverview}. 


\subsection{Keyphrase Vocabulary}
\label{sec:keyphrase_vocab}



CASPER representation units include tokens and keyphrases. For tokens, we follow previous work and use BERT vocabulary. For keyphrases, we build a common and comprehensive keyphrase vocabulary, which we detail in this section.

\noindent \textbf{Scientific corpus}. We start with the S2ORC corpus \cite{lo-etal-2020-s2orc}, which include 81.1M academic papers, spanning across multiple disciplines. From this corpus\footnote{We utilize the pre-processed version provided by sentence-transformers: \url{https://huggingface.co/datasets/sentence-transformers/s2orc}}, we sample 10M papers, using the concatenated title and abstract of each as the source text. We denote the resulting collection as $\mathcal{D} = \{d_1, d_2, ...\}$.

\noindent \textbf{Extracting keyphrases from corpus texts}. To ensure building a vocabulary containing common, widely used keyphrases, we extract them from corpus texts using a  keyphrase extraction algorithm. In particular, we employ ERU-KG \cite{do2025eru}, which is not only effective but also time-efficient (an attribute that is important in our case as there are millions of articles to process). Using the extraction mode of ERU-KG\footnote{setting $\alpha, \beta =1$ as in the original paper}, we extract a set of keyphrases $\mathcal{K}_{d_i}$ for each document $d_i \in \mathcal{D}$.


\noindent \textbf{Forming the vocabulary}. We aim at a keyphrase vocabulary that is comprehensive, so that documents can be effectively represented with these keyphrases. To achieve this, we model vocabulary construction as an instance of the \textit{maximum coverage problem} \cite{nemhauser1978analysis, hochbaum1997approximating}. After obtaining $\mathcal{K}_{d_i}$ for all $d_i \in \mathcal{D}$, we form a \textit{document set} $\mathcal{D}_{k}=\{d_i \in \mathcal{D} \mid k \in \mathcal{K}_{d_i}\}$ for each keyphrase $k$. From the set of all candidate keyphrases, $\mathcal{K}=\bigcup_{d_i\in\mathcal{D}}\mathcal{K}_{d_i}$, our goal is to select a subset $V_k \subset \mathcal{K}$ of predetermined size $|V_k|$ that solves the following optimization problem
\begin{equation}
\label{eq:maximum_coverage_objective}
    \max \quad |\bigcup_{k \in V_k} \mathcal{D}_k|
\end{equation}
where $|V_k|$ is a hyperparameter and $|V_k| \ll |\mathcal{K}|$.

As this problem is NP-hard \cite{nemhauser1978analysis, hochbaum1997approximating}, we approximate the solution using a greedy algorithm \cite{hochbaum1997approximating}. In particular, $V_k$ is constructed by repeatedly adding the keyphrase that appears in the most previously uncovered documents until it reaches the predetermined size.

\noindent \textbf{Integration of the keyphrase vocabulary}. We augment the BERT vocabulary with the keyphrases in $V_k$, treating each as a new token. Because these newly added keyphrase tokens have randomly initialized embeddings and MLM parameters, we further pretrain the modified BERT model, using the MLM training objective, to help it learn effective representations for the keyphrases. We continue to pretrain the BERT model using the scientific article collection $\mathcal{D}$.

\subsection{Pooling Strategy}
\label{sec:pooling_strategy}

SPLADE leverages max pooling to determine the importance of individual vocabulary tokens, which has been empirically shown to outperform sum pooling in retrieval tasks \cite{formal2021spladev2}. Nevertheless, this approach may not fully capture the relevance of research concepts. Max pooling excels at identifying the most salient tokens, since a vocabulary token is included in the representation if it receives a strong activation from any input token. However, this does not necessarily align with how research concepts operate within scientific texts, where the relevance of a concept is often indicated by its consistent significance across a document. With this in mind, we propose a hybrid pooling strategy: max pooling for vocabulary tokens to preserve SPLADE’s benefits, and sum pooling for concepts to better reflect their importance.
\begin{equation}
\resizebox{0.85\columnwidth}{!}{
    $w_j^{d} = 
    \begin{cases}
        \max_{i \in d} \log (1 + \text{ReLU}(w_{ij}^{d})), & \text{if $j \in V_t$}\\
        \sum_{i \in d} \log (1 + \text{ReLU}(w_{ij}^{d})), & \text{if $j \in V_k$}
    \end{cases}$
}
\end{equation}

\subsection{Training CASPER}
\label{sec:training}

Although tokens and keyphrases both act as representation units, they encode different information. Specifically, token-based representations reflect fine-grained distinctions, whereas keyphrase-based representations capture differences in research concepts. This motivates decoupling the training for each representation type, where each type is trained with different negative examples, so that token- and keyphrase-based representations are trained to differentiate on suitable granularity.

Formally, the ranking loss is defined as
\begin{equation}
\resizebox{0.85\columnwidth}{!}{
    $\mathcal{L}_{rank-IBN} = \lambda^t \mathcal{L}^{t}_{rank-IBN} + \lambda^k \mathcal{L}^{k}_{rank-IBN}$
}
\label{eq:CASPER_ranking_loss}
\end{equation}
where $\mathcal{L}^{t}_{rank-IBN}$, $\mathcal{L}^{k}_{rank-IBN}$ are ranking losses for training token and keyphrase-based representation, respectively. $\mathcal{L}^{t}_{rank-IBN}$ is defined the same as in Eq. \ref{eq:SPLADE_ranking_loss}, where the negative documents include both in-batch and hard negatives. In-batch negatives are essentially random negatives and are likely to be different from positive documents in terms of fine-grained details and research concepts. Hard negatives likely share similar research concepts but are different from positive documents in terms of fine-grained details. We define $\mathcal{L}^{k}_{rank-IBN}$ similar to $\mathcal{L}^{t}_{rank-IBN}$, but without the hard negative document in the denominator. The reason is that we want keyphrase-based representation to capture differences in research concepts.

\subsection{Inference}

\label{sec:inference}

At inference time, a document $d$ is ranked based on their similarity with query $q$ in terms of fine-grained details and research concepts. Specifically, documents are ranked based on the following function
\begin{equation}
\label{eq:CASPER_inference_ranking_score}
S(q,d) =s^t(q,d) + \beta \ s^k(q,d) 
\end{equation}
where $\beta$ is a hyperparameter that controls the influence of the concept-level similarity. $s^t(q,d) = \sum_{j \in V_t} w^d_j w^q_j$ and $s^k(q,d) = \sum_{j \in V_k} w^d_j w^q_j$ is the token and concept matching score, respectively.


\subsection{CASPER for Keyphrase Generation}


\label{sec:casper_keyphrase_gen}

Beyond retrieval, CASPER's sparse representations can be adapted for keyphrase generation. By ranking noun phrases extracted from the document against CASPER's sparse weights, we can identify present keyphrases. Simultaneously, highly activated keyphrases $k \in V_k$ in the sparse vector that do not appear in the text serve as absent keyphrases. We provide further details in Appendix \ref{sec:casper_keyphrase_gen_appendix}.

\section{FRIEREN}

\label{sec:frieren}

In this section, we describe FRIEREN, a framework for mining \textbf{FR}ee sc\textbf{IE}ntific \textbf{RE}trieval supervisio\textbf{N}. FRIEREN mines queries from four sources, namely \textit{co-citations}, \textit{citation contexts}, \textit{author-assigned keyphrases} and \textit{titles}. We provide an example of these four sources in Figure \ref{fig:ExampleScholarlyReferences}. Queries mined from these sources are then used to augment user queries, creating a dataset for training. Formally, we aim to build a collection of triplets, denoted as $\mathcal{T}=$ $\{(q_i, d^+_i, d^-_i)\}_{i=1}^{|\mathcal{T}|}$. Here, $q_i$ denotes a query and $d^+_i$, $d^-_i$ denotes the corresponding positive and hard negative document.

\noindent \textbf{User queries}. We utilize SciRepEval Search\footnote{\url{https://huggingface.co/datasets/allenai/scirepeval/viewer/search/train}} \cite{singh-etal-2023-scirepeval}. Each query $q_i$ in this set is associated with a list of candidates and their relevance scores. We select candidates whose scores $\geq 1$ as $d^+_i$. Then, a negative document $d^-_i$ is randomly chosen among those whose scores are 0. 

\begin{table*}[ht]
\centering

\resizebox{0.85\textwidth}{!}{

\begin{tabular}{@{}lccccccccc@{}}
\toprule
\multicolumn{1}{l|}{}                                           & SciFact       & SCIDOCS       & NFCorpus      & DORIS-MAE     & CSFCube       & ACM-CR        & LitSearch     & \multicolumn{1}{c|}{RELISH}        & AVG           \\ \midrule
\multicolumn{10}{c}{\textbf{NDCG@10}}                                                                                                                                                                                                \\ \midrule
\multicolumn{1}{l|}{BM25}                                       & 67.9          & 14.9          & 32.2          & 20.8          & {\ul 5.8}     & 31.2          & 40.2          & \multicolumn{1}{c|}{60.9}          & 34.2          \\
\multicolumn{1}{l|}{SPECTER2}                                   & 63.9          & 17.1          & 24            & 16.3          & 5             & 22.3          & 33.2          & \multicolumn{1}{c|}{59}            & 30.1          \\
\multicolumn{1}{l|}{E5-base-v2}                                 & \textbf{71.4} & \textbf{18.9} & \textbf{35.5} & 16.9          & 5.3           & 28.7          & 40.2          & \multicolumn{1}{c|}{65.3}          & 35.3          \\
\multicolumn{1}{l|}{DyVo}                                       & 68.2          & 15.3          & {\ul 34.5}    & 9.6           & \textbf{6.5}  & 26.6          & 40.5          & \multicolumn{1}{c|}{59.9}          & 32.6          \\
\multicolumn{1}{l|}{ColBERT}                                    & 65.7          & 14.7          & 30.9          & 20.1          & 4.9           & 28.5          & 37.6          & \multicolumn{1}{c|}{60.1}          & 32.8          \\
\multicolumn{1}{l|}{SPLADEv2}                                   & 63.8          & 14.9          & 32            & 17.2          & 5.3           & 27.2          & 42.8          & \multicolumn{1}{c|}{60.6}          & 33            \\
\multicolumn{1}{l|}{ColBERT$_{\text{FRIEREN}}$}                 & 70.4          & 16.8          & 33            & \textbf{23.9} & 5.7           & 34            & 46.2          & \multicolumn{1}{c|}{\textbf{66.9}} & {\ul 37.1}    \\
\multicolumn{1}{l|}{SPLADEv2$_{\text{FRIEREN}}$}                & \textbf{71.4} & 16.8          & 27.1          & 17.8          & 4.6           & 33.8          & 41.6          & \multicolumn{1}{c|}{65.1}          & 34.8          \\
\multicolumn{1}{l|}{SPLADEv2$_{\text{FRIEREN}}^{\text{words}}$} & 70.5          & 16.9          & 28.4          & 20.8          & 5.3           & 32.1          & 43.8          & \multicolumn{1}{c|}{64.7}          & 35.3          \\
\multicolumn{1}{l|}{CASPER}                                     & 70.2          & {\ul 18.8}    & 32.2          & 21.3          & 5.5           & {\ul 34.2}    & {\ul 46.8}    & \multicolumn{1}{c|}{65}            & 36.8          \\
\multicolumn{1}{l|}{CASPER++}                                   & 70.6          & 18.5          & 34.2          & {\ul 23.7}    & 5.6           & \textbf{36.7} & \textbf{49}   & \multicolumn{1}{c|}{{\ul 65.5}}    & \textbf{38}   \\ \midrule
\multicolumn{10}{c}{\textbf{Recall@100}}                                                                                                                                                                                             \\ \midrule
\multicolumn{1}{l|}{BM25}                                       & 92.5          & 34.8          & 24.6          & 34.4          & 16.7          & 55.9          & 74.4          & \multicolumn{1}{c|}{71.8}          & 50.6          \\
\multicolumn{1}{l|}{SPECTER2}                                   & 94.6          & \textbf{45.7} & 25            & 34.4          & \textbf{26.8} & 53.1          & 72            & \multicolumn{1}{c|}{60.3}          & 51.5          \\
\multicolumn{1}{l|}{E5-base-v2}                                 & 95.3          & 42.8          & 31.8          & 35.1          & 22            & 59.7          & 76.4          & \multicolumn{1}{c|}{70}            & 54.1          \\
\multicolumn{1}{l|}{DyVo}                                       & 93.4          & 36.3          & 28.5          & 15.4          & 20.9          & 53.1          & 75.9          & \multicolumn{1}{c|}{66.8}          & 48.8          \\
\multicolumn{1}{l|}{ColBERT}                                    & 88            & 34.2          & 27            & 33.9          & 17.5          & 51.9          & 72.5          & \multicolumn{1}{c|}{63.4}          & 48.6          \\
\multicolumn{1}{l|}{SPLADEv2}                                   & 89.3          & 35            & 26.8          & 35.2          & 17.8          & 54.4          & 75.8          & \multicolumn{1}{c|}{66.9}          & 50.2          \\
\multicolumn{1}{l|}{ColBERT$_{\text{FRIEREN}}$}                 & 95.6          & 41.5          & 29.5          & {\ul 42.2}    & 24.3          & 60.8          & 79.2          & \multicolumn{1}{c|}{{\ul 74}}      & 55.9          \\
\multicolumn{1}{l|}{SPLADEv2$_{\text{FRIEREN}}$}                & 95.7          & 42.5          & 27.2          & 39.3          & 23.2          & 62.7          & 78.6          & \multicolumn{1}{c|}{72.4}          & 55.2          \\
\multicolumn{1}{l|}{SPLADEv2$_{\text{FRIEREN}}^{\text{words}}$} & 95.3          & 41.7          & 30.4          & 40            & 20.7          & 63.6          & 79.1          & \multicolumn{1}{c|}{73}            & 55.5          \\
\multicolumn{1}{l|}{CASPER}                                     & {\ul 95.8}    & {\ul 43.5}    & \textbf{31.9} & 41.6          & {\ul 25.2}    & {\ul 65.6}    & {\ul 82.1}    & \multicolumn{1}{c|}{72.8}          & {\ul 57.3}    \\
\multicolumn{1}{l|}{CASPER++}                                   & \textbf{96.1} & 43            & \textbf{31.9} & \textbf{47.3} & 23.1          & \textbf{66.2} & \textbf{82.4} & \multicolumn{1}{c|}{\textbf{75.1}} & \textbf{58.1} \\ \bottomrule
\end{tabular}

}
\caption{Retrieval performance on eight benchmark datasets. We report performance using nDCG@10 and Recall@100 (in percentage points). The best and second best results are \textbf{bolded} and \uline{underlined}, respectively. We also provide Recall@1000 performance in Table \ref{tab:retrieval_performance_additional_recall1000}.}
\label{tab:retrieval_performance}

\vspace{-0.5cm}
\end{table*}


\noindent \textbf{Co-citations}. Using the full-text S2ORC corpus\footnote{\url{https://github.com/allenai/s2orc}}, we identify citation groups within each paper. From each group, we randomly select two articles: one's title serves as the query $q_i$, and the other's concatenated title and abstract serves as the positive document $d^+_i$. The hard negative $d^-_i$ is an article cited elsewhere in the same paper but outside that specific citation group.

\noindent \textbf{Citation contexts}. The sentence in which a paper is cited often summarizes its key contribution as perceived by other authors. For each full-text document within the S2ORC corpus, we extract the citing sentences treat them as queries $q_i$. The cited paper's title and abstract form the positive document $d^+_i$, while a hard negative $d^-_i$ is another paper cited elsewhere in the same document.

\noindent \textbf{Author-assigned keyphrases}. Keyphrases reflect an author's own view of their work's core concepts. We utilize two keyphrase generation datasets namely KP20K \cite{meng-etal-2017-deep} and KPBioMed \cite{houbre-etal-2022-large}, where documents are annotated with author-assigned keyphrases. For each entry, the comma-separated list of keyphrases serves as $q_i$, the paper itself as the positive document $d^+_i$, and a randomly sampled document from the same dataset acts as the negative $d^-_i$. We note that while random sampling yields easy negatives, we adopt this approach due to the lack of reliable heuristics for identifying hard negatives for this data type.

\noindent \textbf{Titles}. A document's title is a concise, author-provided summary of its content. Again, using the S2ORC corpus, we treat each paper's title as $q_i$ and its corresponding abstract as $d^+_i$. The negative document $d^-_i$ is the abstract of a paper cited within the full text of the source document.

\section{Experiments}

\subsection{Datasets \& Evaluation Metrics}
We utilize a total of eight scientific retrieval datasets. Among them, three are sourced from the BEIR benchmarks \cite{thakur2021beir}, namely SciFact \cite{wadden-etal-2020-fact}, SCIDOCS \cite{cohan-etal-2020-specter} and NFCorpus \cite{boteva2016full}. The other five include DORIS-MAE \cite{wang2023scientific}, CSFCube \cite{mysore2021csfcube}, ACM-CR \cite{boudin2021acm}, LitSearch \cite{ajith2024litsearch} and RELISH \cite{brown2019large}. Table \ref{tab:eval_dataset_statistics} summarizes the statistics of the evaluation datasets used in our experiments. To measure retrieval effectiveness, we employ nDCG@10 and Recall@100. We select nDCG@10 as the standard metric for ranking quality, while Recall@100 is included to evaluate the model's suitability as a first-stage retriever, as rerankers typically process the top-100 documents \cite{thakur2021beir}. Unless noted, results represent a single run.




\subsection{Retrieval Performance Evaluation}
\label{sec:retrieval_performance_evaluation}

\noindent \textbf{Baselines}. To ensure a fair comparison, we select models with comparable parameter counts namely SPECTERv2 \cite{singh-etal-2023-scirepeval}, E5-base-v2 \cite{wang2022text}, DyVo \cite{nguyen2024dyvo}, SPLADEv2-max \cite{formal2021spladev2}, and ColBERT \cite{khattab2020colbert}, along with BM25. Additionally, we introduce CASPER++, a simple ensemble of CASPER and BM25 obtained by summing their scores, following the approach of \cite{formal2022distillation}. 

We also include versions of SPLADEv2-max and ColBERT finetuned on the FRIEREN dataset, denoted as SPLADEv2${_{\text{FRIEREN}}}$ and ColBERT${_{\text{FRIEREN}}}$, respectively. This setup serves two main purposes: 1) to assess the contribution of FRIEREN by comparing these models to their original versions; and 2) to ensure CASPER is evaluated against baselines trained under identical conditions. For both models, we employ similar hyperparameters as in the original papers, modifying only the training batch size to 20 and maximum sequence length to 256 to match CASPER's configurations. We provide further details regarding implementations of baseline methods in \S \ref{sec:baseline_implementation_details}.

\noindent \textbf{Comparison with baselines}. The performance of our proposed method and the baselines are presented in Table \ref{tab:retrieval_performance}. When compared to baselines not trained with FRIEREN, CASPER achieves better performance in terms of both nDCG@10 and Recall@100. Notably, CASPER surpasses E5-base-v2, despite the latter being trained on a much larger dataset. 



CASPER demonstrates good performance when compared to baselines trained with FRIEREN. In terms of nDCG@10, CASPER outperforms SPLADEv2${_{\text{FRIEREN}}}$, but slightly falls behind ColBERTv2${_{\text{FRIEREN}}}$. In terms of Recall@100, CASPER outperforms both models. Overall, this demonstrates that even when trained on identical data,  CASPER surpasses other strong baselines, indicating that the concept-integrated representation we introduce provides a tangible advantage.

To verify that performance gains are driven by the concept-integrated representation rather than just the expanded vocabulary, we conducted an additional comparison. In particular, CASPER was evaluated against SPLADEv2${^{\text{words}}_{\text{FRIEREN}}}$, a version of SPLADEv2${_{\text{FRIEREN}}}$ that included an extra vocabulary of 30k most common English words in $\mathcal{D}$ but not present in BERT's token vocabulary. This enhanced SPLADEv2 underperforms relative to CASPER, which further supports the effectiveness of the proposed concept-integrated representation.

\noindent \textbf{FRIEREN generates suitable training data for scientific document search}. Comparing baselines trained with FRIEREN to their original counterparts, we find that models trained with FRIEREN generally achieve competitive nDCG@10 and higher Recall@100. Overall, this shows that training data generated by FRIEREN is indeed beneficial for training scientific text retrievers.

\noindent \textbf{Effectiveness of CASPER++}. Finally, examining the performance of CASPER++, we observe that it surpasses CASPER, and also all evaluated baselines. This improvement demonstrates that integrating BM25 continues to provide complementary benefits to CASPER, which is consistent with the observation made in \cite{formal2022distillation}. Notably, for this integration, we tokenize inputs using CASPER’s tokenizer, and therefore allow BM25 to operate on sequences of both tokens and keyphrases.

\begin{figure}[ht]
    \centering
    \includegraphics[width=\columnwidth]{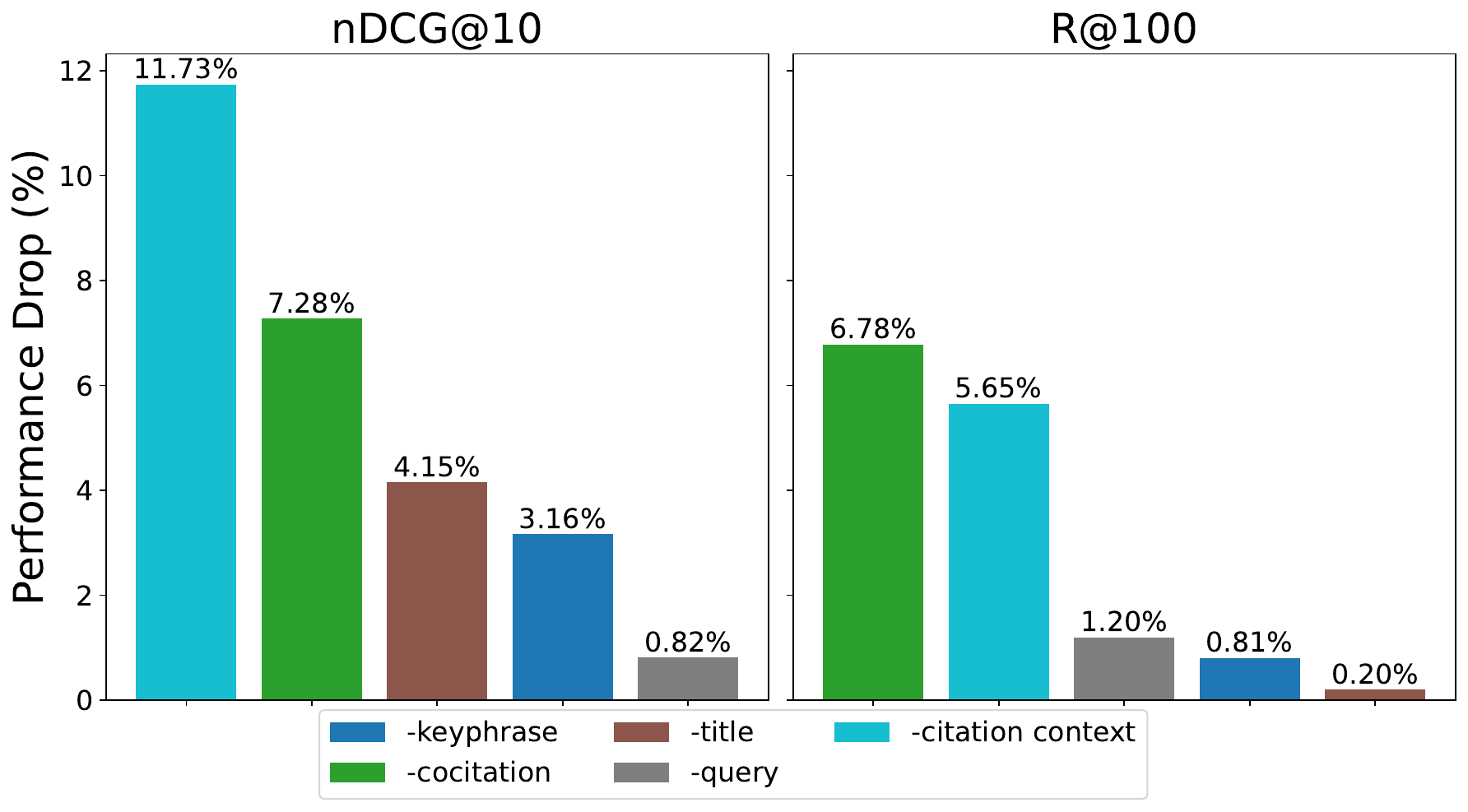} 
    \caption{Retrieval performance drop as different data sources are removed.}
    \label{fig:data_sources_ablation_study}
\end{figure}

\begin{figure*}[h]
\centering
\includegraphics[width=0.85\textwidth]{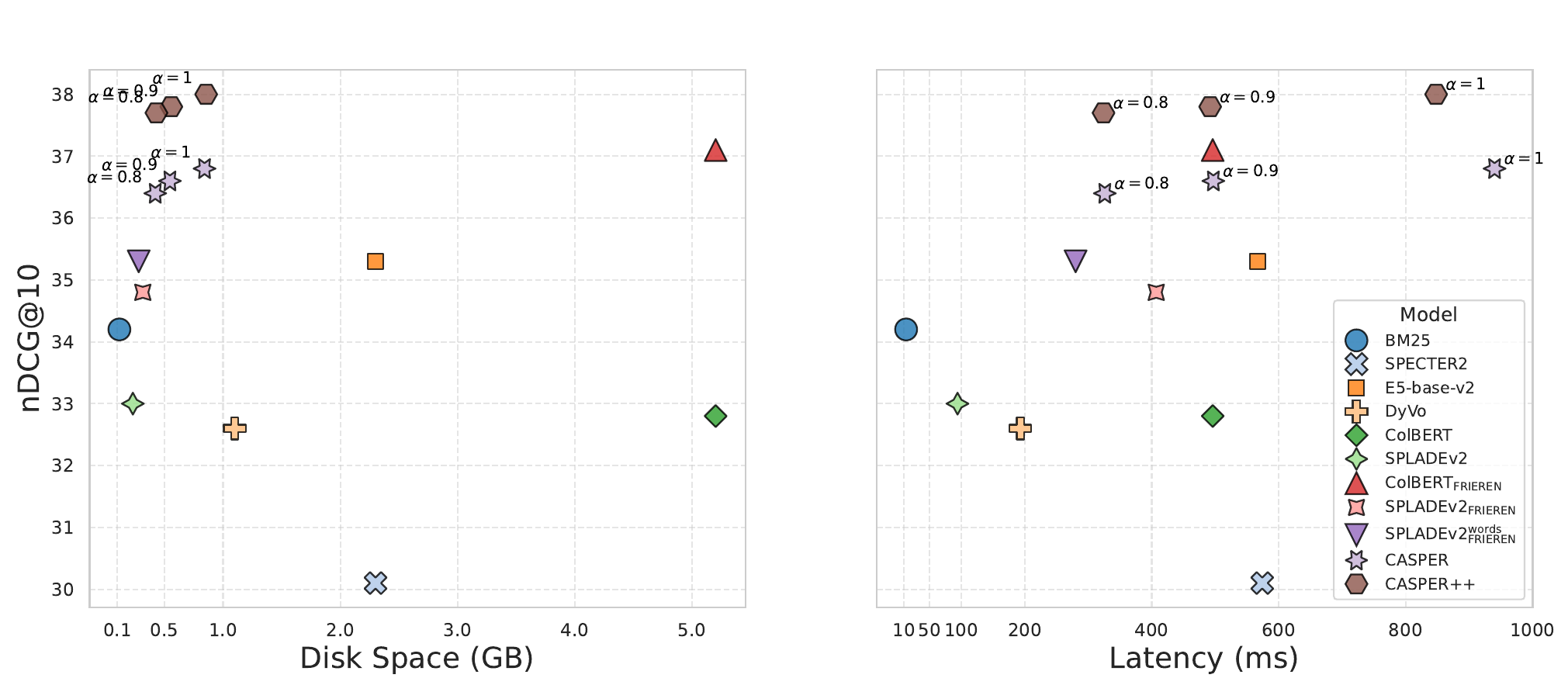}

\caption{Trade-off between effectiveness (nDCG@10) and efficiency (Disk Space and Latency) for CASPER, CASPER++ (at varying pruning levels), and baselines. nDCG@10 is averaged across eight benchmark datasets. Efficiency metrics (Disk Space in GB; Latency in ms) are measured on CSFCube, the largest dataset, using a single-threaded CPU. Latency values exclude encoding time and are averaged over three runs.}
\label{fig:performance_efficiency}
\end{figure*}

\subsection{Keyphrase Generation Performance}

We evaluate CASPER on five standard benchmarks using the KPEval framework \cite{wu2023kpeval}. Compared to CopyRNN \cite{meng-etal-2017-deep}, a strong seq2seq baseline, CASPER falls slightly behind in semantic F1. However, it generates more diverse keyphrases and is more than three times faster. A full experimental setup, quantitative results, and a qualitative case study comparing the two models are detailed in Appendix \ref{sec:casper_keyphrase_gen_appendix}.

\subsection{Impact of Data Sources}
\label{sec:impact_of_data_sources}
In this section, we aim to understand the contribution of each data source within the FRIEREN framework to CASPER’s overall performance. Specifically, we train five versions of CASPER, each by removing one type of data from the full training set. Results are presented in Figure \ref{fig:data_sources_ablation_study}, which shows that all data sources contribute positively to the final model, as omitting any single type leads to a decrease in performance.


Among the data types, \textit{user queries} is the data type that results in insignificant performance drops across both evaluation metrics, indicating that the contribution of user queries is limited. This observation further supports our argument regarding user queries and interaction data being an inadequate source of supervision for training retrievers in the scientific domain. In contrast, citation contexts and co-citations emerge as the most important sources of information, since removing either results in a substantial reduction in performance. Notably, both citation contexts and co-citations are readily available and can be scaled up with ease, unlike user queries, which are more challenging to collect and contribute less to the final outcome.






\subsection{Efficiency}

We investigate the impact of pruning CASPER's query and document representations on both effectiveness and efficiency. To prune a sparse vector $w$, we retain its $\alpha$-mass subvector, following the definition proposed by \citet{bruch2024efficient}. We conduct experiments with $\alpha \in \{0.8, 0.9, 1.0\}$, where $\alpha = 1.0$ represents the default CASPER.

Figure \ref{fig:performance_efficiency} illustrates the trade-off between retrieval effectiveness and efficiency across varying pruning levels, alongside competitive baselines. Regarding storage efficiency, we observe that sparse modeling baselines are superior to their dense counterparts, requiring significantly less disk space. In terms of latency, most models are comparable, clustering within the 200 to 600ms range. While the default CASPER ($\alpha=1.0$) exhibits higher latency, the pruned variants ($\alpha \in \{0.8, 0.9\}$) achieve competitive inference speeds without compromising retrieval performance.

\subsection{Ablation Studies}

We summarize our key ablation studies in this section. First, in \S \ref{sec:influence_of_concept_based_representation}, we examine the impact of the concept-level score by varying $\beta$ (see \S \ref{sec:inference}); our results indicate that concept-based matching yields optimal performance when it complements, rather than dominates, token-based matching. Second, in \S \ref{sec:ablation_studies_keyphrase_vocab}, we analyze the effect of keyphrase vocabulary variations. We find that an intermediate vocabulary size is most effective; and that ensuring comprehensiveness during vocabulary construction (see \S \ref{sec:keyphrase_vocab}) is necessary. Finally, supporting our discussion in \S \ref{sec:pooling_strategy}, we empirically demonstrate in \S \ref{sec:sum_vs_max_pooling} that sum pooling outperforms max pooling for generating concept representations. 


\section{Conclusion}


In this paper, we present CASPER, a sparse model for scientific document retrieval designed to represent queries and documents by their research concepts and match them at both granular and conceptual levels. We also propose FRIEREN, a framework that mines data from scholarly references to overcome the lack of suitable supervision signals for scientific document search. Through extensive experiments, we show that CASPER generally outperforms strong baselines across eight scientific retrieval benchmarks. We also explore the effectiveness-efficiency trade-off via representation pruning and demonstrate CASPER's interpretability by showing its utility in the task of keyphrase generation.



\section*{Limitations}



We acknowledge several limitations in our current work. \textbf{First}, we did not incorporate knowledge distillation techniques, as explored in prior studies \cite{formal2021spladev2, santhanam-etal-2022-colbertv2}. Applying distillation from strong cross-encoders or LLMs could likely further enhance the performance of CASPER and other baselines trained on FRIEREN. \textbf{Second}, computational constraints limited the scale of our experiments; although FRIEREN is capable of generating extensive training data, we have not yet explored the effects of scaling up to larger models and datasets. We leave these investigations to future work.

\textbf{Finally}, CASPER operates on a fixed keyphrase vocabulary, requiring retraining to incorporate new or modified concepts. In contrast, approaches like DyVo \cite{nguyen2024dyvo} offer greater flexibility by retrieving entities and scoring them based on their descriptions. However, DyVo relies on a frozen retriever and lacks end-to-end optimization. A promising direction for future research is to bridge this gap, combining CASPER’s effective end-to-end learning with the dynamic flexibility of description-based models.

\bibliography{custom}

\appendix
\label{sec:appendix}

\section{CASPER for Keyphrase Generation}

\label{sec:casper_keyphrase_gen_appendix}

In this section, we show that through simple post-processing, CASPER can be used for the keyphrase generation task, which aims to identify present and absent keyphrases given a text $d$. Keyphrases generated by CASPER can be used to interpret its representations.

\noindent \textbf{Present keyphrases}. This type of keyphrases can be found within the input text. To identify present keyphrases, we first extract noun phrases from $d$ to form a candidate set $C=\{c_i\}_i$. Next, we obtain sparse representation of $d$ produced by CASPER $w^d=\{w^d_j\}_{j \in V_t \cup V_k}$. Then, highest-ranked candidates are chosen to be keyphrases, where the ranking function is defined as
\begin{equation}
R(c_i) = \frac{\omega_d(c_i)}{|c_i| - \gamma}\sum_{j \in c_i} w^d_j
\end{equation}

\begin{equation}
    \omega_d(c_i) = 1 + \frac{1}{\log_2(\mathcal{P}_d(c_i) + 2)}
\end{equation}

Here, $\gamma$ and $\omega_d(c_i)$ are inspired by \cite{do2025eru}. The former is a hyperparameter that controls the preference towards longer candidates, while the latter is applied to favor candidates that appear earlier in the text. $\mathcal{P}_d(c_i)$ is the offset position, computed by the number of words that precedes $c_i$ in $d$. We note that the candidate $c_i$ is tokenized over the combined vocabulary $V_t \cup V_k$. For example, ``unsupervised machine learning'', can be tokenized into (``un'', ``\#\#su'', ``\#\#per'', ``\#\#vis'', ``\#\#ed'', ``machine learning'').

\noindent \textbf{Absent keyphrases}. Simultaneously, absent keyphrases are identified by selecting the keyphrases from $V_k$ that are not in the document but possess the highest activation weights $w^d_k$ in the sparse vector. A key advantage of this method is its exceptional speed. Specifically, by not relying on sequence-to-sequence generation \cite{meng-etal-2017-deep} or external phrase retrieval for absent keyphrases \cite{do2025eru}, CASPER offers a time-efficient generation capability.

\section{Implementation Details}
\label{sec:implementation_details}

\subsection{CASPER}

We initialize CASPER using DistilBERT$_{\text{base}}$. Unless otherwise specified, the keyphrase vocabulary size is set to $|V_k| = $ 30k. After incorporating keyphrases into the base model (see \S\ref{sec:keyphrase_vocab}), we conduct continual pretraining with a batch size of 64, a learning rate of $2 \times 10^{-5}$, and a total of 70k steps. To prioritize effective learning of new keyphrase tokens, we modify the masking strategy: within each sequence, 85\% of keyphrase tokens are selected for prediction, while the rest of the masking procedure follows the original BERT methodology.

Following pretraining, we finetune CASPER using training data generated by FRIEREN (\S\ref{sec:frieren}). To generate the dataset, we use only the first 7\% of the S2ORC corpus\footnote{\url{https://github.com/allenai/s2orc}} (20 files out of a total of 298 files). We sample a maximum of 1.5 million triplets per data type to manage resource usage and prevent over-representation of larger categories. As some types are smaller (e.g., \textit{title} with 330,000 triplets), the resulting dataset yields 3.6 million triplets.

Hyperparameters for finetuning largely follows the settings described in \cite{formal2021splade}, with some adjustments. Specifically, we use a batch size of 20 due to resource limitations. For all experiments, we set $\lambda_q = 3 \times 10^{-4}$ and $\lambda_d = 1 \times 10^{-4}$ (see Eq. \ref{eq:SPLADE_training_objective}), while for the ranking loss, we set $\lambda_t = 1$ and $\lambda_k = 2$ (see Eq. \ref{eq:CASPER_ranking_loss}). During inference, unless otherwise specified, we use $\beta=0.25$.



\subsection{Baselines Implementation Details}

\label{sec:baseline_implementation_details}

We utilize the public pretrained models of SPECTERv2\footnote{\url{https://huggingface.co/allenai/specter2_base}}, E5-base-v2\footnote{\url{https://huggingface.co/intfloat/e5-base-v2}} and SPLADEv2-max\footnote{\url{https://huggingface.co/naver/splade_v2_max}}. Our setup differs only in that we initialize with DistilBERT-base, use a batch size of 20, and train on MS MARCO using BM25 hard negatives, consistent with the SPLADE training protocol \cite{formal2021splade, formal2021spladev2}. During inference, we employ the PLAID engine \cite{santhanam2022plaid} for both ColBERT and ColBERT$_{\text{FRIEREN}}$. Finally, for DyVo \cite{nguyen2024dyvo}, we train the model using the official repository and report results for the DyVo-BM25 variant.

\subsection{Processing CSFCube}

For the CSFCube dataset, which provides relevance judgments for three aspects (“background,” “method,” and “result”), we combine these judgments into a single dataset (with 127, 73, and 81 judgments for each aspect, respectively). Queries with relevance judgments for multiple aspects are treated as distinct queries.

\subsection{Computational Resources}

We conduct all our experiments on a server with NVIDIA Ampere A40 GPUs (300W, 48GB VRAM each), along with two AMD EPYC 7302 3GHz CPUs and 256 GB of RAM. We employ 1 GPU at a time for all experiments.

\begin{table}[]
\centering

\resizebox{0.8\columnwidth}{!}{

\begin{tabular}{@{}lccc@{}}
\toprule
\multicolumn{4}{c}{\textbf{Information Retrieval}}                 \\ \midrule
\multicolumn{1}{c|}{Dataset name} & \#Query & \#Corpus  & Avg D /Q \\ \midrule
\multicolumn{1}{l|}{SciFact}      & 300     & 5,183     & 1.1      \\
\multicolumn{1}{l|}{SCIDOCS}      & 1,000   & 25,657    & 4.9      \\
\multicolumn{1}{l|}{NFCorpus}     & 323     & 3,633     & 38.2     \\
\multicolumn{1}{l|}{DORIS-MAE}    & 100     & 363,133   & 16.4     \\
\multicolumn{1}{l|}{CSFCube}      & 50      & 776,070   & 5.64     \\
\multicolumn{1}{l|}{ACM-CR}       & 552     & 114,882   & 1.8      \\
\multicolumn{1}{l|}{LitSearch}    & 597     & 64,183    & 1.1      \\
\multicolumn{1}{l|}{RELISH}       & 1,684   & 163,170   & 80       \\ \midrule
\multicolumn{4}{c}{\textbf{Keyphrase Generation}}                  \\ \midrule
\multicolumn{1}{c|}{Dataset name} & \#doc   & \#kps/doc & \%absent \\ \midrule
\multicolumn{1}{l|}{SemEval}      & 100     & 15.2      & 59.7     \\
\multicolumn{1}{l|}{Inspec}       & 500     & 9.8       & 22       \\
\multicolumn{1}{l|}{NUS}          & 211     & 11.6      & 49.3     \\
\multicolumn{1}{l|}{Krapivin}     & 460     & 5.7       & 51.2     \\
\multicolumn{1}{l|}{KP20K}        & 19,987  & 5.3       & 44.7     \\ \bottomrule
\end{tabular}

}
\caption{Statistics of evaluation datasets}
\label{tab:eval_dataset_statistics}

\end{table}

\begin{table*}[ht]
\centering

\resizebox{0.9\textwidth}{!}{

\begin{tabular}{@{}lccccccccc@{}}
\toprule
\multicolumn{1}{l|}{}                                           & SciFact      & SCIDOCS       & NFCorpus      & DORIS-MAE     & CSFCube       & ACM-CR        & LitSearch     & \multicolumn{1}{c|}{RELISH}      & AVG           \\ \midrule
\multicolumn{10}{c}{\textbf{Recall@1000}}                                                                                                                                                                                         \\ \midrule
\multicolumn{1}{l|}{BM25}                                       & 97.7         & 56.4          & 37            & {\ul 77.1}    & 46.4          & 71.5          & 90            & \multicolumn{1}{c|}{93.4}        & 71.2          \\
\multicolumn{1}{l|}{SPECTER2}                                   & 99.7         & \textbf{77.8} & 59.7          & 70.9          & 53.8          & 76.1          & 87.1          & \multicolumn{1}{c|}{90}          & 76.9          \\
\multicolumn{1}{l|}{E5-base-v2}                                 & 99.7         & 70            & \textbf{64.7} & 69.6          & 54.4          & 77.6          & 91.4          & \multicolumn{1}{c|}{93.6}        & 77.6          \\
\multicolumn{1}{l|}{DyVo}                                       & 99.0         & 59.3          & 52.4          & 38.5          & 44.6          & 72.1          & 90.6          & \multicolumn{1}{c|}{90.6}        & 68.4          \\
\multicolumn{1}{l|}{ColBERT}                                    & 96           & 57            & 45.3          & 65.9          & 44.8          & 70.4          & 87.6          & \multicolumn{1}{c|}{88.2}        & 69.4          \\
\multicolumn{1}{l|}{SPLADEv2}                                   & 97.7         & 57            & 52.2          & 65.6          & 43            & 73.4          & 92.8          & \multicolumn{1}{c|}{91}          & 71.6          \\
\multicolumn{1}{l|}{ColBERTv2$_{\text{FRIEREN}}$}               & 99.3         & 66.5          & 56.1          & 77            & 52.7          & 81.2          & 91.6          & \multicolumn{1}{c|}{94.8}        & 77.4          \\
\multicolumn{1}{l|}{SPLADEv2$_{\text{FRIEREN}}$}                & 99.3         & 68.4          & 57.7          & 76.9          & 54.6          & 79.6          & 93            & \multicolumn{1}{c|}{94.8}        & 78            \\
\multicolumn{1}{l|}{SPLADEv2$_{\text{FRIEREN}}^{\text{words}}$} & 98.7         & 68.8          & 62.1          & 75.7          & 54.7          & 80.6          & 93.3          & \multicolumn{1}{c|}{95.2}        & 78.6          \\
\multicolumn{1}{l|}{CASPER}                                     & \textbf{100} & {\ul 70.6}    & 63.1          & {\ul 77.1}    & \textbf{60.2} & {\ul 81.3}    & \textbf{94.9} & \multicolumn{1}{c|}{{\ul 95.4}}  & {\ul 80.3}    \\
\multicolumn{1}{l|}{CASPER++}                                   & \textbf{100} & 69.3          & {\ul 63.5}    & \textbf{81.5} & {\ul 57.9}    & \textbf{82.6} & {\ul 94.8}    & \multicolumn{1}{c|}{\textbf{96}} & \textbf{80.7} \\ \bottomrule
\end{tabular}

}
\caption{Recall@1000 of baselines on eight retrieval benchmarks. The best and second-best results are \textbf{bolded} and \underline{underlined}.}
\label{tab:retrieval_performance_additional_recall1000}

\end{table*}

\section{Keyphrase Generation Evaluation}

\begin{table}[]
\centering

\resizebox{\columnwidth}{!}{
\begin{tabular}{@{}l|ccc|ccc@{}}
\toprule
\multirow{2}{*}{} & \multicolumn{3}{c|}{\textbf{CopyRNN}}                                         & \multicolumn{3}{c}{\textbf{CASPER}}                                           \\ \cmidrule(l){2-7} 
                  & SemF1$\uparrow$ & SemR$\uparrow$ & \begin{tabular}[c]{@{}c@{}}Diversity\\ (emb\_sim$\downarrow$)\end{tabular} & SemF1$\uparrow$ & SemR$\uparrow$ & \begin{tabular}[c]{@{}c@{}}Diversity\\ (emb\_sim$\downarrow$)\end{tabular} \\ \midrule
Inspec            & 55.0  & 52.5 & 23.8                                                           & 52.6  & 53.7 & 13.0                                                           \\
NUS               & 57.8  & 55.0 & 28.3                                                           & 49.3  & 50.9 & 13.1                                                           \\
Krapivin          & 54.2  & 57.2 & 24.3                                                           & 42.2  & 49.7 & 10.9                                                           \\
SemEval           & 50.4  & 43.0 & 24.8                                                           & 46.7  & 42.1 & 12.5                                                           \\
KP20K             & 54.1  & 57.2 & 26.0                                                           & 43.9  & 51.7 & 11.9                                                           \\
AVG               & 54.3  & 53.0 & 25.4                                                           & 47.0  & 49.6 & 12.3                                                           \\ \midrule
Throughput$\uparrow$        & \multicolumn{3}{c|}{11 (doc/s)}                                               & \multicolumn{3}{c}{36.3 (doc/s)}                                              \\ \bottomrule
\end{tabular}
}

\caption{Keyphrase generation performance on five benchmark datasets, with SemF1, SemR, and emb\_sim reported in percentage points. $\uparrow$ indicates higher is better, and $\downarrow$ indicates otherwise. Throughput, measured in documents per second (doc/s), is averaged over five runs}
\label{tab:keyphrase_gen_performance_and_throughput}

\end{table}

\label{sec:interpretability_evaluation}

To evaluate CASPER's keyphrase generation performance, we utilize five widely used keyphrase generation datasets: SemEval \cite{kim2013automatic}, Inspec \cite{hulth2003improved}, NUS \cite{nguyen2007keyphrase}, Krapivin \cite{krapivin2009large}, and KP20K \cite{meng-etal-2017-deep}. For evaluation metrics, we adopt semantic F1 (SemF1), semantic recall (SemRecall), and diversity (emb\_sim), as provided by the KPEval evaluation framework \cite{wu2023kpeval}. In our evaluation, predictions are generated by selecting the top 5 present and top 5 absent keyphrases for each instance, resulting in a set of 10 predicted keyphrases per document. 

Table \ref{tab:keyphrase_gen_performance_and_throughput} presents CASPER's performance on the keyphrase generation task. For comparison, we evaluate against CopyRNN \cite{meng-etal-2017-deep}, a well-established seq2seq keyphrase generation model, using the implementation from \cite{do-etal-2023-unsupervised}. CASPER attains 86\% of CopyRNN's semantic F1 score, reflecting weaker agreement with ground-truth keyphrases. However, CASPER achieves 94\% of CopyRNN’s semantic recall and generates substantially more diverse keyphrases, suggesting that precision is the more major contribution of CopyRNN’s higher F1, which may be partly due to generating repetitive or highly similar phrases. Although not being able to outperform CopyRNN in terms of groundtruth agreement, CASPER offers significant practical advantages, as it is more than three times faster and generates more diverse keyphrases.
\begin{table}[]

\centering

\resizebox{\columnwidth}{!}{








\begin{tabular}{p{0.3\textwidth} p{0.6\textwidth}} 
\toprule
\multicolumn{1}{c}{\textbf{Document}} & \multicolumn{1}{c}{\textbf{Keyphrases}} \\
\midrule

\textbf{Title}: Adaptive road detection via context-aware label transfer.

\par\vspace{10pt} 

\textbf{Abstract}: The vision ability is fundamentally important for a mobile robot. Many aspects have been investigated during the past few years, but there still remain questions to be answered. This work mainly focuses on the task of road detection, ... 
& 
\textbf{CopyRNN}: \textcolor{customdarkorange}{nearest neighbor search; mobile robot; label transfer; depth map; nearest neighbor;} \textcolor{customdarkblue}{mobile robot vision; context-aware robot; road recovery computing; road recovery}

\par\vspace{10pt} 

\textbf{CASPER}: \textcolor{customdarkorange}{depth map; adaptive road detection; road detection; context-aware label transfer; robot;} \textcolor{customdarkblue}{intelligent vehicle; detection algorithm; computer vision; adaptivity; transfer process}

\par\vspace{10pt} 

\textbf{Targets}: \textcolor{customdarkorange}{road detection; depth map; label transfer; context-aware; mrf; } \textcolor{customdarkblue}{computer vision} \\
\bottomrule
\end{tabular}
}

\caption{Keyphrases generated for an example document, by CopyRNN and CASPER. Highlighted in \textcolor{customdarkorange}{orange} and \textcolor{customdarkblue}{blue} are present and absent keyphrases, respectively}
\label{tab:interpretability_case_studies}
\end{table}

To supplement the automatic metrics, Table \ref{tab:interpretability_case_studies} presents the keyphrases generated by each model for a sample document. Both systems generate reasonable present keyphrases, but they differ when it comes to absent keyphrases. CopyRNN mainly forms absent phrases by recombining existing terms, whereas CASPER proposes truly novel phrases, one of which (“computer vision”) matches the ground-truth annotation. In addition, CopyRNN tends to generate keyphrases that are similar (``nearest neighbor search'' and ``nearest neighbor''; ``mobile robot vision'' and ``robot vision''), which supports our analysis above.



\begin{figure}
    \centering
    \includegraphics[width=\columnwidth]{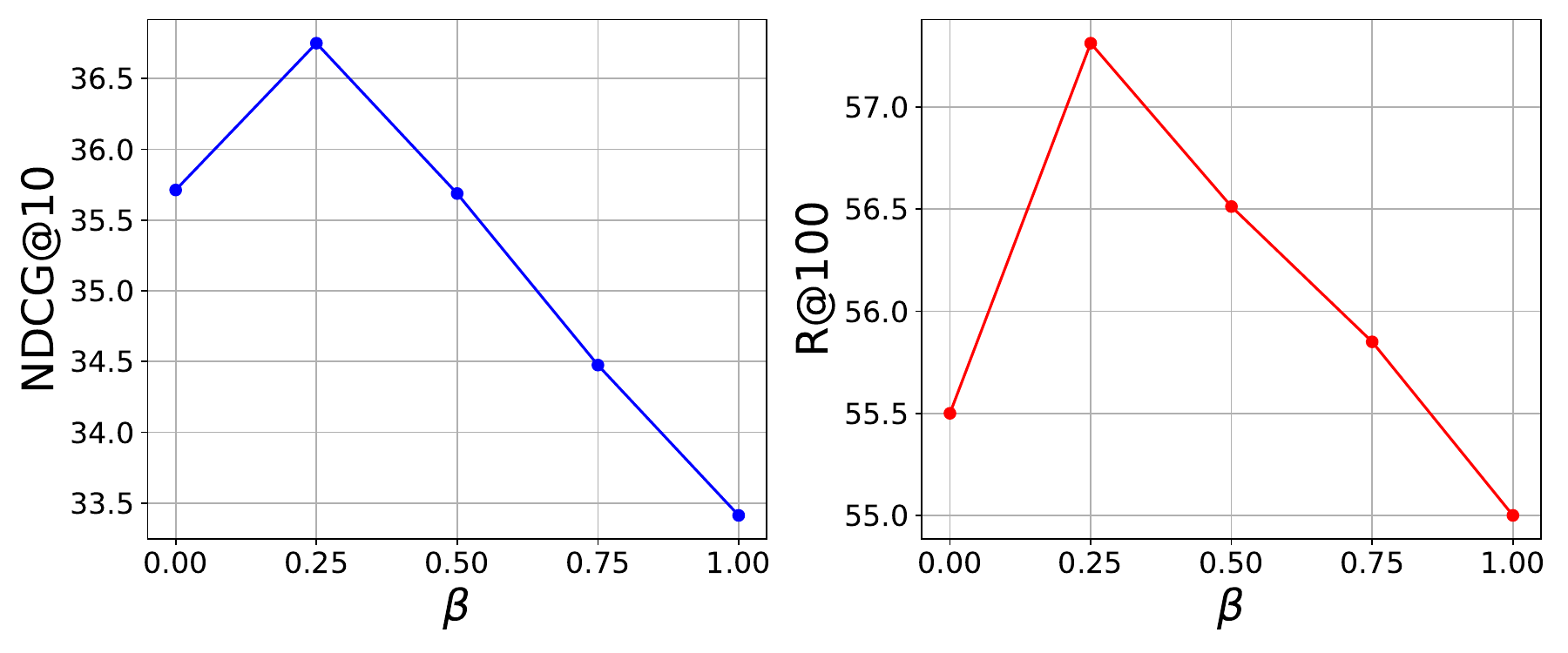}
    \caption{Retrieval performance (averaged across eight datasets) with different values of $\beta$}
    \label{fig:beta_adjustment}
\end{figure}

\section{Ablation Studies}

\subsection{Influence of Concept-based Representation}
\label{sec:influence_of_concept_based_representation}

As described in \S \ref{sec:inference}, the final ranking score in CASPER is computed as a combination of token-level and keyphrase-level (concept-level) scores. To assess the impact of the concept-level score on overall performance, we vary the weighting parameter $\beta$ (see Eq. \ref{eq:CASPER_inference_ranking_score}) across five values: $\{0, 0.25, 0.5, 0.75, 1\}$.

The results, illustrated in Figure \ref{fig:beta_adjustment}, reveal that CASPER achieves its highest performance at $\beta = 0.25$. Notably, when concept-based matching is not used ($\beta = 0$), the performance of CASPER drops significantly. Conversely, a $\beta$ with value beyond 0.25 also results in a continuous decline in performance as it is increased. These findings indicate that, while concept-based matching provides clear benefits, it should complement rather than dominate token-based matching for optimal results.

\begin{table}[]
\centering
\resizebox{0.75\columnwidth}{!}{

\begin{tabular}{@{}cc|cc@{}}
\toprule
\multicolumn{2}{c|}{\textbf{Sum}} & \multicolumn{2}{c}{\textbf{Max}} \\ \midrule
NDCG@10          & R@100          & NDCG@10          & R@100         \\ \midrule
36.8             & 57.3           & 35.7             & 56.2          \\ \bottomrule
\end{tabular}

}
\caption{Average retrieval performance across eight datasets when sum and max pooling are employed for concept-based representation}
\label{tab:sum_vs_max_pooling}

\end{table}

\subsection{Sum vs Max Pooling}
\label{sec:sum_vs_max_pooling}

To support our discussion in \S \ref{sec:pooling_strategy} on the effect of apply sum versus max pooling for forming concept-based representation, we train a version of CASPER where max pooling is used instead of sum pooling. The results, presented in Table \ref{tab:sum_vs_max_pooling}, show that the default version which uses sum pooling achieves better performance.

\begin{table}[]
\centering
\resizebox{0.75\columnwidth}{!}{
\begin{tabular}{@{}lcc@{}}
\toprule
\multicolumn{1}{c|}{Variation}                                                                 & NDCG@10 & R@100 \\ \midrule
\multicolumn{3}{c}{\textbf{Main model}}                                                                          \\ \midrule
\multicolumn{1}{l|}{$|V_k|$ = 30k}                                                             & 36.8    & 57.3  \\ \midrule
\multicolumn{3}{c}{\textbf{Varying vocabulary size}}                                                             \\ \midrule
\multicolumn{1}{l|}{$|V_k|$ = 5k}                                                              & 36.2    & 56.3  \\
\multicolumn{1}{l|}{$|V_k|$= 15k}                                                              & 37.1    & 57.2  \\
\multicolumn{1}{l|}{$|V_k|$ = 60k}                                                             & 35.6    & 55.8  \\ \midrule
\multicolumn{3}{c}{\textbf{Frequency based keyphrase vocabulary}}                                                \\ \midrule
\multicolumn{1}{l|}{\begin{tabular}[c]{@{}l@{}}Frequency-based\\ keyphrase vocab\end{tabular}} & 36.2    & 56.8  \\ \bottomrule
\end{tabular}
}

\caption{Retrieval performance of CASPER with different keyphrase vocabularies. We report the average retrieval performance across eight datasets}
\label{tab:keyphrase_vocab_ablation}

\end{table}

\subsection{Ablation Studies of Keyphrase Vocabulary}
\label{sec:ablation_studies_keyphrase_vocab}

It is important to understand the impact of different keyphrase vocabularies to the final outcome of the model. In this section, we investigate: 1) how varying the size of the keyphrase vocabulary affects effectiveness; and 2) whether explicitly ensuring comprehensiveness is essential in constructing the vocabulary, or if simply selecting the most frequent keyphrases suffices (since a vocabulary including the most frequent keyphrases is also comprehensive to a certain extent). To explore this, we train three additional versions of CASPER with $|V_k| = $ 5k, 15k and 60k. Furthermore, we train another version of CASPER whose vocabulary are formed by selecting the most frequent 30k keyphrases.

The results, presented in Table \ref{tab:keyphrase_vocab_ablation}, show that CASPER achieves the best performance with the default vocabulary setting. Firstly, analyzing performance as we vary vocabulary size, intermediate vocabulary sizes ($|V_k| = $ 15k and 30k) yield better results than either very small ($|V_k| = $ 5k) or very large ($|V_k| = $ 60k) vocabularies. The results suggest that the best performance is achieved when the keyphrase vocabulary is both common and comprehensive. In particular, a vocabulary that is too small is not comprehensive and therefore may lack descriptive power (as it does not allow representing key concepts of many documents). On the other hand, one that is too large introduces specific, low-frequency keyphrases, which not only make it harder to train the model but also offer limited benefit regarding the enhancement of the representation space, as they produce matches fewer documents.

Secondly, the variant using the 30k most frequent keyphrases exhibits similar recall to the default but lower nDCG. This result suggests that there are benefits in explicitly optimizing comprehensiveness when building keyphrase vocabulary.

\subsection{Multi versus Single Disciplinary}
\label{sec:multi_vs_single}

In this section, we evaluate CASPER's performance under two scenarios: single-disciplinary and multi-disciplinary training. Specifically, we investigate the performance of CASPER when tailored to a single discipline (specifically Computer Science in this study) versus when it is trained across multiple scientific fields.

\begin{table}[]
\centering

\resizebox{0.7\columnwidth}{!}{

\begin{tabular}{@{}c|c|cc@{}}
\toprule
\textbf{Mode}           & \textbf{\#Training data} & \textbf{NDCG@10} & \textbf{R@100} \\ \midrule
\multirow{3}{*}{single} & 600k                     & 25.1             & 50.8           \\
                        & 1.5M                     & 24.9             & 53.3           \\
                        & 2.9M                     & 25.4             & 52.2           \\ \midrule
\multirow{3}{*}{multi}  & 600k                     & 24.2             & 50             \\
                        & 1.5M                     & 25.2             & 52.5           \\
                        & 3.6M                     & 27               & 53.6           \\ \bottomrule
\end{tabular}

}
\caption{CASPER performance in multi and single disciplinary setting. We report average performance across the four Computer Science retrieval benchmarks, namely DORIS-MAE, CSFCube, LitSearch  and ACM-CR}
\label{tab:multi_vs_single}


\end{table}

To test this, we create a Computer Science-specific version of CASPER following the same procedure as the default model, with two key differences: 1) the keyphrase vocabulary $V_k$ is constructed using only the Computer Science subset of $\mathcal{D}$, and continuous pretraining is performed also on this subset; 2) using FRIEREN, we generate a Computer Science training dataset by retaining only triplets where the positive document belongs to the Computer Science domain\footnote{To determine if a document belongs to the Computer Science domain, we utilize the ``fieldsOfStudy'' field returned by Semantic Scholar API}. This results in a training set of comparable size (2.9M) to the one used to train the default model (3.6M).

We evaluate both versions across three training data scales: 600k, 1.5M triplets, and full (3.6M for multi-disciplinary and 2.9M for single-disciplinary). For fair comparison, we conduct experiments on four retrieval benchmarks whose main theme is Computer Science, namely DORIS-MAE, CSFCube, LitSearch  and ACM-CR. The results, presented in Table \ref{tab:multi_vs_single}, reveal an interesting pattern. When trained on 600k triplets, the single-disciplinary CASPER consistently outperforms its multi-disciplinary counterpart. However, as training data increases, the multi-disciplinary version improves and eventually achieves slightly superior performance when using 3M triplets. This suggests that while domain-specific training provides advantages with limited data, the multi-disciplinary approach benefits more from increased training scale. We attribute this to cross-domain knowledge transfer, where knowledge from different domains complement one another to achieve better results.

\end{document}